\documentclass[11pt,a4paper]{article}
\usepackage[left=15mm, top=10mm, right=15mm, bottom=20mm, nohead=true, nofoot=true]{geometry}

\usepackage[utf8]{inputenc}
\usepackage{authblk}
\usepackage{indentfirst}
\usepackage{misccorr}
\usepackage{graphicx}
\usepackage{amsmath}
\usepackage{amssymb}
\usepackage{multicol}
\usepackage{bm}
\usepackage{longtable}
\usepackage{pdflscape}
\usepackage{epstopdf}
\usepackage{multirow}
\usepackage{adjustbox}
\usepackage{amsfonts}
\usepackage{ifthen}
\usepackage{fancyhdr}
\usepackage{setspace}
\usepackage{xcolor}
\usepackage[round,authoryear,comma,sort&compress,numbers]{natbib}
\usepackage[toc,title,page]{appendix}
\usepackage[hidelinks]{hyperref}
\usepackage{url}

\newcommand{\pa}{\partial}

\hypersetup{
    colorlinks=true,
    linkcolor=green,
    citecolor=blue,
    filecolor=magenta,
    urlcolor=cyan,
    pdftitle={Sharelatex Example},
    pdfpagemode=FullScreen,
}

\fancypagestyle{plain}{\chead{\texttt{\textcolor{brown}{Communications of BAO, Vol. ?, Issue ?, 20??, pp. ???-???}}}}
\title{\textbf{Unraveling the Origins and Development of the Galactic Disk through  Metal-Poor Stars}}
\author[1,2]{Maria Rah \thanks{E-mail: \href{mailto:mariarah.astro@gmail.com}{mariarah.astro@gmail.com}, Corresponding author}} 
\author[3]{Manolya Yatman}
\author[4]{Ali Taani}
\author[5]{Ahmad A.\ Abushattal}
\author[6]{Mohammad K.\ Mardini}

\affil[1]{\scriptsize Byurakan Astrophysical Observatory, Armenia; R-PhD Student at BAO}
\affil[2]{\scriptsize National Astronomical Observatory, Chinese Academy of Sciences, China; The Silk Road Project at NAOC}
\affil[3]{\scriptsize Physics Department, Bryn Mawr College, 19010, Philadelphia, USA}
\affil[4]{\scriptsize Physics Department, Faculty of Science, Al Balqa Applied University,19117, Salt, Jordan}
\affil[5]{\scriptsize Department of Physics, Al-Hussein Bin Talal University, P.O. Box 20, 71111, Ma’an, Jordan}
\affil[6]{\scriptsize Department of Physics, Zarqa University, Zarqa 13110, Jordan}

\def\actaa{Acta Astronomica. }
\def\apj{Astrophys.~J. }

\def\aap{Astron.~Astrophys. }

\def\apjl{Astrophys.~J.~Lett. }
\def\apjs{Astrophys.~J.~Suppl.~Ser. }
\def\aj{Astron.~J. }
\def\mnras{Mon.~Not.~R.~Astron.~Soc. }

\begin{document}
\pagestyle{empty}
\newpage
\pagestyle{fancy}
\label{firstpage}
\date{}
\maketitle

\begin{abstract}
The Milky Way is a spiral galaxy comprising three main components: the Bulge, the Disk, and the Halo. Of particular interest is the Galactic disk, which holds a significant portion of the baryonic matter angular momentum and harbors at least two primary stellar populations: the thin and thick disks. Understanding the formation and evolution of the Galactic disk is crucial for comprehending the origins and development of our Galaxy. Stellar archaeology offers a means to probe the disk's evolution by listening to the cosmological narratives of its oldest and most pristine stars, specifically the metal-poor stars. In this study, we employed accurate photometric metallicity estimates and Gaia Early Data Release 3 astrometry to curate a pure sample of the oldest Galactic stars. This proceeding presents a summary of our primary findings. 
\end{abstract}

%https://combao.bao.am/instructionsauthors.php

\section{Introduction}
The pioneering research carried out by \citet{gilmore83} centered on investigating the stellar populations present in our Milky Way galaxy. The study extensively examined the motion and distribution of metals among stars, yielding valuable insights into the structure and evolution of our galaxy. An important discovery stemming from this research was the identification of distinct stellar groups, categorized according to their motion characteristics. These stars were divided into two main clusters: a thin disk population characterized by low velocity dispersion, and a thick disk population exhibiting higher velocity dispersion. This finding implies that diverse dynamic processes contributed to the formation and development of these two components. Furthermore, as stated by \citet{gilmore83}, both the thin and thick disk populations exhibited various metallicities; however, younger stars within the thin disk component generally displayed a preference for higher metallicity levels. This supports theories that suggest star formation in the Milky Way took place over a long period of time, and that subsequent generations of stars showed higher levels of enrichment with heavy elements. We have collected the key characteristics of these two distinct populations, as reported in literature and listed in Table~\ref{tab:populations}.

\begin{table*}[ht!]
	\begin{center}
		\centering
		\caption{Orbital properties of the Galactic thin disk, thick disk, and inner halo}
		\label{tab:populations}
		\begin{tabular}{lccccr}
			\hline
			\hline	
			 Parameter & unit & Thin disk &  Thick disk & Inner halo & Atari disk\\
			\hline	
		     $h_{R}$&   (kpc)  &2.6 - 3.00 & 2.0 - 3.0 & ... &  2.48 $\pm$ 0.05 \\
			 $h_{Z}$&   (kpc) &0.14 - 0.36 & 0.5 - 1.1 & ... & 1.68$^{+0.19}_{-0.15}$ \\
			 $<V_{\phi}>$&   (km s$^{-1}$)               &  208 & 182 &   0 &  154 $\pm$ 1\\
		     Z$_{max}$&   (kpc)                         & $< 0.8$   & $0.8$ - $3.0$ & $> 3.0$  &  $<$ 3.0\\
		     e&  ....         & $< 0.14$   & 0.3 - 0.5 & $>$ 0.7 & 0.30 - 0.7\\
			\hline
		\end{tabular}
  \\
See \citet{Mardini2022}; and references therein.
	\end{center}
\end{table*}

More recently, \citet{Carollo2019} identified two distinct modes of the Galactic thick-disk: ``in-situ'' and ``ex-situ''. The in-situ population refers to the formation of stars within the Galactic disk, while the ex-situ population involves stars that were formed elsewhere and later accreted onto the our Galaxy. This component is referred to as the Metal-Weak Thick Disk. In this context, \citet{Mardini2022} investigated the nature of this component by employing accurate photometric metallicity estimates and Gaia Early Data Release 3 astrometries \citep{Gaiadr3}. Furthermore, during the Gaia era, numerous studies have utilized these astrometries to yield essential constraints regarding the formation and evolution of our galaxy, the Milky Way \citep{2017A&A...606A..45D, Chiti_map,Chiti_SMSS_cat2021,Mardini2022b,Dhaim2022,Zepeda2023,Mardini23_dwarf,Hong2023,Placco2023}. In a nutshell, the study by \citet{Mardini2022b} offers significant insights into the formation and evolution of our Galaxy by emphasizing the spatial, kinematic, and chemical characteristics of this component, which suggest that multiple mechanisms contribute to Galactic disk growth.

\section{Method and Analysis}

We have developed two separate techniques, namely velocity and action space analysis, to initially select a pure sample of stars with [Fe/H] $\leq -0.8$ form the APOGEE-2/SDSS-IV dataset \citep{Blanton}. By displaying kinematics that are characteristic of the thick disk, our resulting sample consists of 90,000 stars with high-quality measurements of [Fe/H] and astrometry. We then calculated the positions of our stars using the equations described in Equations~\ref{eq:1}. The velocity calculations were carried out according to Equation~\ref{eq:2}. In order to categorize our sample into the main Galactic componentrelative s (i.e., Halo, thin disk, and thick disk), we defined velocity distributions as explained in Equation~\ref{eq:3}. The relative probabilities for the ratios between thick-disk-to-thin-disk (TD/D) and thick-disk-to-halo (TD/H) were determined using Equation~\ref{eq:4}. Each thick disk star was assigned with a membership probability of TD/D $>$ 2.0, while stars with TD/D $<$ 0.5 were classified as thin disk stars. Moreover, we excluded stars with TD/H $<$ 10.0 to minimize potential contamination from the Galactic halo.

\begin{figure}
\centering
\includegraphics[width=\textwidth]{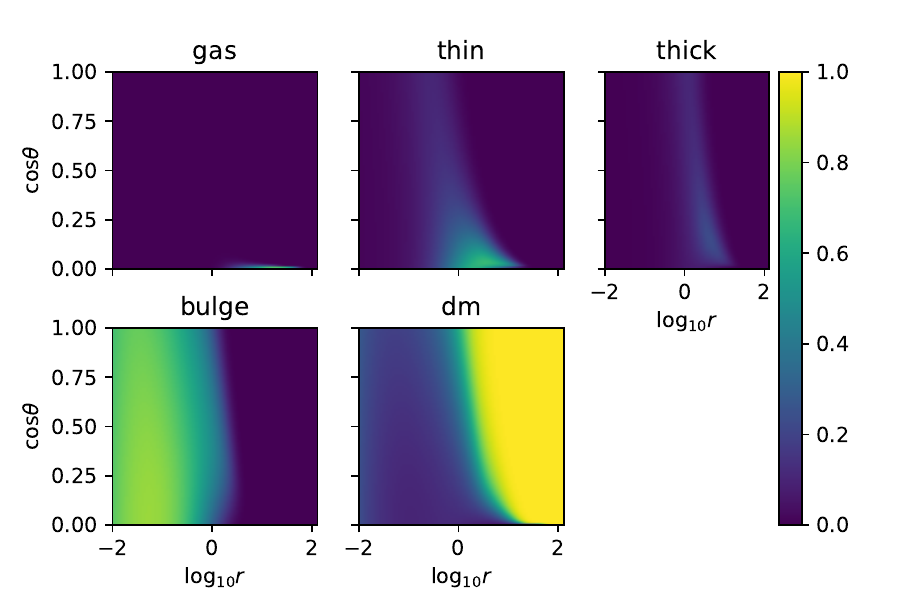}
\caption{The relative density of each of the components.}
\label{fig:components}
\end{figure}

The above-mentioned method efficiently identifies stars exhibiting disk-like kinematics. Nonetheless, it is crucial to acknowledge that this approach might erroneously categorize halo stars, which have nearly-circular orbits and low orbital eccentricities, as disk-like stars \citep[e.g.,][]{Mardini_2019a,Mardini_2019b,Mardini_2020,Placco2020,Brauer2022}.
To minimize the risk of contamination from the Galactic Halo, we have developed an alternative method that relies on stellar actions, as described by Equation~\ref{eq:5}. Furthermore, we utilized a spherically symmetric ad hoc approximation to evaluate the three distribution functions (DFs) individually, as outlined in Equation~\ref{eq:6}. The figure depicting the relative density distribution of each Galactic component, along with the dark matter profile, can be found in Figure~\ref{fig:components}. It is worth mentioning that the potential demonstrates a primarily isotropic characteristic \cite[see][]{Taani2019,Taani2019b,Taani2020,Almusleh2021,Mardini2019c,Taani2019JPhCS,Taani2022JHEAp}. This yielded a sample of my 87075 stars, which spans a wide metallicity range.

\section{Results and Conclusions}

The average rotational velocity of our sample shows a 30\,km\,s$^{-1}$ lag compared to the well-known thick disk. To understand the origins of our sample stars, we analyze derived gradients, the shape of the eccentricity distribution, and theoretical scenarios for thick disk formation \citep{Masda2019,Abdusalam2020,Tawalbeh2021,Wardat2021}. We also calculate the scale height and scale length of our sample using Equations~\ref{eq:boltzmann}, \ref{eq:Jeans2}, \ref{eq:Jeans3}, and \ref{eq:Jeans4}. In terms of size, our sample is similar to the Galactic thick disk in the radial direction but has greater vertical extension. Furthermore, the distribution of orbital eccentricities in our sample bridges between those typically observed in the thick disk and halo populations. Our investigation into orbital eccentricities also reveals a significant number of stars with high eccentricities. These findings, combined with theoretical predictions, suggest that this population was introduced to our Galaxy through an early merger event involving the proto-Milky Way.

\begin{equation}
\begin{split}\label{eq:1}
X= R_{\odot}-d\cos{(l)} \cos{(b)}
\\
Y= -d \sin{(l)}\cos{(b)}
\\
Z=d\sin{(b)}
\end{split}
\end{equation}

\begin{equation} \label{eq:2}
    \begin{split}
      V_R = U \cos{\phi}+(V+V_{rot}) \sin{\phi}\\
      V_\phi =(V+V_{rot})\cos{\phi}- U \sin{\phi}\\
      V_z=W  
    \end{split}
\end{equation}

\begin{align} \label{eq:3}
f(U,V,W)=k\cdot \textrm{exp}(&-\frac{(V_{\textrm{LSR}}-V_{\textrm{asym}})^2}{2\sigma_{V}^2}\nonumber\\
&-\frac{W_{\textrm{LSR}}^2}{2\sigma_{W}^2} -\frac{U_{\textrm{LSR}}^2}{2\sigma_{U}^2})
\end{align}

\begin{equation} \label{eq:4}
    \begin{split}
TD/D=\frac{X_{TD}.f_{TD}}{X_D . f_D}\\
TD/H=\frac{X_{TD}.f_{TD}}{X_H.f_H}
    \end{split}
\end{equation}

\begin{equation}\label{eq:5}
    f_{halo}(J_r,J_z,L_z)=f_0\left[1+\frac{J_r+J_z+|L_z|}{J_0}\right]^{\beta_*}
\end{equation}

\begin{equation}\label{eq:6}
    \Phi_{\mathrm{approx}}(r)= -\Phi_{0,\mathrm{fit}} \frac{r_{\mathrm{fit}}}{r}\left[1-\frac{1}{(1+r/r_{\mathrm{fit}})^{\beta_{\mathrm{fit}}-3}}\right]
\end{equation}

\begin{align} \label{eq:boltzmann}
\frac{\pa f}{\pa t}  	& + v_R \frac{\pa f}{\pa R} + \frac{v_\phi}{R^{2}} \frac{\pa f}{\pa \phi} + v_z \frac{\pa f}{\pa z} - \left(\frac{\pa \Phi}{\pa R} - \frac{v_\phi^2}{R^{3} } \right) \frac{\pa f}{\pa v_R} \nonumber \\
			  	& - \frac{\pa \Phi}{\pa\phi} \frac{\pa f}{\pa v_\phi} - \frac{\pa \Phi}{\pa z}\frac{\pa f}{\pa v_z} = 0,
\end{align}

\begin{align} \label{eq:Jeans2}
\rho K_{Z}= \frac{\pa (\rho\sigma^{2}_{V_{Z}})}{\pa Z} + \frac{1}{R} \frac{\pa (R\rho\sigma^{2}_{V_{R,Z}})}{\pa R}
\end{align}

\begin{align} \label{eq:Jeans3}
\frac{\sigma^{2}_{V_{\phi}}}{\sigma^{2}_{V_{R}}}-2+\frac{2R}{h_{R}}-\frac{V_{c}^{2} -\bar{V_{\phi}}^{2} }{\sigma^{2}_{V_{R}}}+\frac{\sigma^{2}_{V_{Z}}}{\sigma^{2}_{V_{R}}} = 0
\end{align}

\begin{align} \label{eq:Jeans4}
\frac{\pa \ln{\sigma^{2}_{V_{Z}}}}{\pa Z} - \frac{1}{h_{Z}}  + \frac{K_{Z}}{\sigma^{2}_{V_{Z}}}=0
\end{align}

% \clearpage % To force this stuff to happen by this point in the text, otherwise these will probably end up after the references.

\scriptsize
\bibliographystyle{ComBAO}
\nocite{*}
\bibliography{references}

\begin{thebibliography}{}
\makeatletter
\relax
\def\mn@urlcharsother{\let\do\@makeother \do\$\do\&\do\#\do\^\do\_\do\%\do\~}
\def\mn@doi{\begingroup\mn@urlcharsother \@ifnextchar [ {\mn@doi@}
  {\mn@doi@[]}}
\def\mn@doi@[#1]#2{\def\@tempa{#1}\ifx\@tempa\@empty \href
  {http://dx.doi.org/#2} {doi:#2}\else \href {http://dx.doi.org/#2} {#1}\fi
  \endgroup}
\def\mn@eprint#1#2{\mn@eprint@#1:#2::\@nil}
\def\mn@eprint@arXiv#1{\href {http://arxiv.org/abs/#1} {{\tt arXiv:#1}}}
\def\mn@eprint@dblp#1{\href {http://dblp.uni-trier.de/rec/bibtex/#1.xml}
  {dblp:#1}}
\def\mn@eprint@#1:#2:#3:#4\@nil{\def\@tempa {#1}\def\@tempb {#2}\def\@tempc
  {#3}\ifx \@tempc \@empty \let \@tempc \@tempb \let \@tempb \@tempa \fi \ifx
  \@tempb \@empty \def\@tempb {arXiv}\fi \@ifundefined
  {mn@eprint@\@tempb}{\@tempb:\@tempc}{\expandafter \expandafter \csname
  mn@eprint@\@tempb\endcsname \expandafter{\@tempc}}}

\bibitem[\protect\citeauthoryear{{Abdusalam}, {Ablimit}, {Hashim}, {L{\"u}},
  {Mardini}  \& {Wang}}{{Abdusalam} et~al.}{2020}]{Abdusalam2020}
{Abdusalam} K.,  {Ablimit} I.,  {Hashim} P.,  {L{\"u}} G.~L.,  {Mardini} M.~K.,
    {Wang} Z.~J.,  2020, \mn@doi [\apj] {10.3847/1538-4357/abb5a8}, \href
  {https://ui.adsabs.harvard.edu/abs/2020ApJ...902..125A} {902, 125}

\bibitem[\protect\citeauthoryear{{Abu-Dhaim}, {Taani}, {Tanineah}, {Tamimi},
  {Mardini}  \& {Al-Wardat}}{{Abu-Dhaim} et~al.}{2022}]{Dhaim2022}
{Abu-Dhaim} A.,  {Taani} A.,  {Tanineah} D.,  {Tamimi} N.,  {Mardini} M.,
  {Al-Wardat} M.,  2022, \mn@doi [\actaa] {10.32023/0001-5237/72.3.2}, \href
  {https://ui.adsabs.harvard.edu/abs/2022AcA....72..171A} {72, 171}

\bibitem[\protect\citeauthoryear{{Al-Tawalbeh} et~al.,}{{Al-Tawalbeh}
  et~al.}{2021}]{Tawalbeh2021}
{Al-Tawalbeh} Y.~M.,  et~al., 2021, \mn@doi [Astrophysical Bulletin]
  {10.1134/S199034132101003X}, \href
  {https://ui.adsabs.harvard.edu/abs/2021AstBu..76...71A} {76, 71}

\bibitem[\protect\citeauthoryear{{Al-Wardat} et~al.,}{{Al-Wardat}
  et~al.}{2021}]{Wardat2021}
{Al-Wardat} M.~A.,  et~al., 2021, \mn@doi [Research in Astronomy and
  Astrophysics] {10.1088/1674-4527/21/7/161}, \href
  {https://ui.adsabs.harvard.edu/abs/2021RAA....21..161A} {21, 161}

\bibitem[\protect\citeauthoryear{{Almusleh}, {Taani}, {{\"O}zdemir}, {Rah},
  {Al-Wardat}, {Zhao}  \& {Mardini}}{{Almusleh} et~al.}{2021}]{Almusleh2021}
{Almusleh} N.~A.,  {Taani} A.,  {{\"O}zdemir} S.,  {Rah} M.,  {Al-Wardat}
  M.~A.,  {Zhao} G.,   {Mardini} M.~K.,  2021, \mn@doi [Astronomische
  Nachrichten] {10.1002/asna.202113867}, \href
  {https://ui.adsabs.harvard.edu/abs/2021AN....342..625A} {342, 625}

\bibitem[\protect\citeauthoryear{{Blanton} et~al.,}{{Blanton}
  et~al.}{2017}]{Blanton}
{Blanton} M.~R.,  et~al., 2017, \mn@doi [\aj] {10.3847/1538-3881/aa7567}, \href
  {https://ui.adsabs.harvard.edu/abs/2017AJ....154...28B} {154, 28}

\bibitem[\protect\citeauthoryear{{Brauer}, {Andales}, {Ji}, {Frebel},
  {Mardini}, {G{\'o}mez}  \& {O'Shea}}{{Brauer} et~al.}{2022}]{Brauer2022}
{Brauer} K.,  {Andales} H.~D.,  {Ji} A.~P.,  {Frebel} A.,  {Mardini} M.~K.,
  {G{\'o}mez} F.~A.,   {O'Shea} B.~W.,  2022, \mn@doi [\apj]
  {10.3847/1538-4357/ac85b9}, \href
  {https://ui.adsabs.harvard.edu/abs/2022ApJ...937...14B} {937, 14}

\bibitem[\protect\citeauthoryear{{Carollo} et~al.,}{{Carollo}
  et~al.}{2019}]{Carollo2019}
{Carollo} D.,  et~al., 2019, \mn@doi [\apj] {10.3847/1538-4357/ab517c}, \href
  {https://ui.adsabs.harvard.edu/abs/2019ApJ...887...22C} {887, 22}

\bibitem[\protect\citeauthoryear{{Chiti}, {Frebel}, {Mardini}, {Daniel}, {Ou}
  \& {Uvarova}}{{Chiti} et~al.}{2021a}]{Chiti_SMSS_cat2021}
{Chiti} A.,  {Frebel} A.,  {Mardini} M.~K.,  {Daniel} T.~W.,  {Ou} X.,
  {Uvarova} A.~V.,  2021a, \mn@doi [\apjs] {10.3847/1538-4365/abf73d}, \href
  {https://ui.adsabs.harvard.edu/abs/2021ApJS..254...31C} {254, 31}

\bibitem[\protect\citeauthoryear{{Chiti}, {Mardini}, {Frebel}  \&
  {Daniel}}{{Chiti} et~al.}{2021b}]{Chiti_map}
{Chiti} A.,  {Mardini} M.~K.,  {Frebel} A.,   {Daniel} T.,  2021b, \mn@doi
  [\apjl] {10.3847/2041-8213/abd629}, \href
  {https://ui.adsabs.harvard.edu/abs/2021ApJ...911L..23C} {911, L23}

\bibitem[\protect\citeauthoryear{{Chiti} et~al.,}{{Chiti}
  et~al.}{2023}]{Chiti2022}
{Chiti} A.,  et~al., 2023, \mn@doi [\aj] {10.3847/1538-3881/aca416}, \href
  {https://ui.adsabs.harvard.edu/abs/2023AJ....165...55C} {165, 55}

\bibitem[\protect\citeauthoryear{{Dai}, {Szkody}, {Taani}, {Garnavich}  \&
  {Kennedy}}{{Dai} et~al.}{2017}]{2017A&A...606A..45D}
{Dai} Z.,  {Szkody} P.,  {Taani} A.,  {Garnavich} P.~M.,   {Kennedy} M.,  2017,
  \mn@doi [\aap] {10.1051/0004-6361/201731310}, \href
  {https://ui.adsabs.harvard.edu/abs/2017A&A...606A..45D} {606, A45}

\bibitem[\protect\citeauthoryear{{Gaia Collaboration} et~al.,}{{Gaia
  Collaboration} et~al.}{2023}]{Gaiadr3}
{Gaia Collaboration} et~al., 2023, \mn@doi [\aap]
  {10.1051/0004-6361/202243940}, \href
  {https://ui.adsabs.harvard.edu/abs/2023A&A...674A...1G} {674, A1}

\bibitem[\protect\citeauthoryear{{Gilmore} \& {Reid}}{{Gilmore} \&
  {Reid}}{1983}]{gilmore83}
{Gilmore} G.,  {Reid} N.,  1983, mnras, \href
  {http://adsabs.harvard.edu/abs/1983MNRAS.202.1025G} {202, 1025}

\bibitem[\protect\citeauthoryear{{Hong} et~al.,}{{Hong}
  et~al.}{2023}]{Hong2023}
{Hong} J.,  et~al., 2023, \mn@doi [arXiv e-prints] {10.48550/arXiv.2311.02297},
  \href {https://ui.adsabs.harvard.edu/abs/2023arXiv231102297H} {p.
  arXiv:2311.02297}

\bibitem[\protect\citeauthoryear{Mardini et~al.,}{Mardini
  et~al.}{2019a}]{Mardini_2019a}
Mardini M.~K.,  et~al., 2019a, \mn@doi [The Astrophysical Journal]
  {10.3847/1538-4357/ab0fa2}, 875, 89

\bibitem[\protect\citeauthoryear{Mardini, Placco, Taani, Li  \& Zhao}{Mardini
  et~al.}{2019b}]{Mardini_2019b}
Mardini M.~K.,  Placco V.~M.,  Taani A.,  Li H.,   Zhao G.,  2019b, \mn@doi
  [The Astrophysical Journal] {10.3847/1538-4357/ab3047}, 882, 27

\bibitem[\protect\citeauthoryear{{Mardini}, {Ershiadat}, {Al-Wardat}, {Taani},
  {{\"O}zdemir}, {Al-Naimiy}  \& {Khasawneh}}{{Mardini}
  et~al.}{2019c}]{Mardini2019c}
{Mardini} M.~K.,  {Ershiadat} N.,  {Al-Wardat} M.~A.,  {Taani} A.~A.,
  {{\"O}zdemir} S.,  {Al-Naimiy} H.,   {Khasawneh} A.,  2019c, in Journal of
  Physics Conference Series. p. 012024 (\mn@eprint {arXiv} {1904.09608}),
  \mn@doi{10.1088/1742-6596/1258/1/012024}

\bibitem[\protect\citeauthoryear{Mardini et~al.,}{Mardini
  et~al.}{2020}]{Mardini_2020}
Mardini M.~K.,  et~al., 2020, \mn@doi [The Astrophysical Journal]
  {10.3847/1538-4357/abbc13}, 903, 88

\bibitem[\protect\citeauthoryear{{Mardini} et~al.,}{{Mardini}
  et~al.}{2022a}]{Mardini2022b}
{Mardini} M.~K.,  et~al., 2022a, \mn@doi [\mnras] {10.1093/mnras/stac2783},
  \href {https://ui.adsabs.harvard.edu/abs/2022MNRAS.517.3993M} {517, 3993}

\bibitem[\protect\citeauthoryear{{Mardini}, {Frebel}, {Chiti}, {Meiron},
  {Brauer}  \& {Ou}}{{Mardini} et~al.}{2022b}]{Mardini2022}
{Mardini} M.~K.,  {Frebel} A.,  {Chiti} A.,  {Meiron} Y.,  {Brauer} K.~V.,
  {Ou} X.,  2022b, \mn@doi [\apj] {10.3847/1538-4357/ac8102}, \href
  {https://ui.adsabs.harvard.edu/abs/2022ApJ...936...78M} {936, 78}

\bibitem[\protect\citeauthoryear{{Mardini}, {Frebel}, {Betre}, {Jacobson},
  {Norris}  \& {Christlieb}}{{Mardini} et~al.}{2023}]{Mardini23_dwarf}
{Mardini} M.~K.,  {Frebel} A.,  {Betre} L.,  {Jacobson} H.,  {Norris} J.~E.,
  {Christlieb} N.,  2023, \mn@doi [arXiv e-prints] {10.48550/arXiv.2305.05363},
  \href {https://ui.adsabs.harvard.edu/abs/2023arXiv230505363M} {p.
  arXiv:2305.05363}

\bibitem[\protect\citeauthoryear{{Masda}, {Docobo}, {Hussein}, {Mardini},
  {Al-Ameryeen}, {Campo}, {Khan}  \& {Pathan}}{{Masda}
  et~al.}{2019}]{Masda2019}
{Masda} S.~G.,  {Docobo} J.~A.,  {Hussein} A.~M.,  {Mardini} M.~K.,
  {Al-Ameryeen} H.~A.,  {Campo} P.~P.,  {Khan} A.~R.,   {Pathan} J.~M.,  2019,
  \mn@doi [Astrophysical Bulletin] {10.1134/S1990341319040126}, \href
  {https://ui.adsabs.harvard.edu/abs/2019AstBu..74..464M} {74, 464}

\bibitem[\protect\citeauthoryear{{Placco} et~al.,}{{Placco}
  et~al.}{2020}]{Placco2020}
{Placco} V.~M.,  et~al., 2020, \mn@doi [\apj] {10.3847/1538-4357/ab99c6}, \href
  {https://ui.adsabs.harvard.edu/abs/2020ApJ...897...78P} {897, 78}

\bibitem[\protect\citeauthoryear{{Placco} et~al.,}{{Placco}
  et~al.}{2023}]{Placco2023}
{Placco} V.~M.,  et~al., 2023, \mn@doi [arXiv e-prints]
  {10.48550/arXiv.2310.17024}, \href
  {https://ui.adsabs.harvard.edu/abs/2023arXiv231017024P} {p. arXiv:2310.17024}

\bibitem[\protect\citeauthoryear{{Taani}, {Karino}, {Song}, {Al-Wardat},
  {Khasawneh}  \& {Mardini}}{{Taani} et~al.}{2019a}]{Taani2019}
{Taani} A.,  {Karino} S.,  {Song} L.,  {Al-Wardat} M.,  {Khasawneh} A.,
  {Mardini} M.~K.,  2019a, \mn@doi [Research in Astronomy and Astrophysics]
  {10.1088/1674-4527/19/1/12}, \href
  {https://ui.adsabs.harvard.edu/abs/2019RAA....19...12T} {19, 012}

\bibitem[\protect\citeauthoryear{{Taani}, {Abushattal}  \& {Mardini}}{{Taani}
  et~al.}{2019b}]{Taani2019b}
{Taani} A.,  {Abushattal} A.,   {Mardini} M.~K.,  2019b, \mn@doi [Astronomische
  Nachrichten] {10.1002/asna.201913713}, \href
  {https://ui.adsabs.harvard.edu/abs/2019AN....340..847T} {340, 847}

\bibitem[\protect\citeauthoryear{{Taani}, {Karino}, {Song}, {Mardini},
  {Al-Wardat}, {Abushattal}, {Khasawneh}  \& {Al-Naimiy}}{{Taani}
  et~al.}{2019c}]{Taani2019JPhCS}
{Taani} A.,  {Karino} S.,  {Song} L.,  {Mardini} M.,  {Al-Wardat} M.,
  {Abushattal} A.,  {Khasawneh} A.,   {Al-Naimiy} H.,  2019c, in Journal of
  Physics Conference Series. p. 012029,
  \mn@doi{10.1088/1742-6596/1258/1/012029}

\bibitem[\protect\citeauthoryear{{Taani}, {Khasawneh}, {Mardini}, {Abushattal}
  \& {Al-Wardat}}{{Taani} et~al.}{2020}]{Taani2020}
{Taani} A.,  {Khasawneh} A.,  {Mardini} M.,  {Abushattal} A.,   {Al-Wardat} M.,
   2020, arXiv e-prints, \href
  {https://ui.adsabs.harvard.edu/abs/2020arXiv200203011T} {p. arXiv:2002.03011}

\bibitem[\protect\citeauthoryear{{Taani}, {Vallejo}  \& {Abu-Saleem}}{{Taani}
  et~al.}{2022}]{Taani2022JHEAp}
{Taani} A.,  {Vallejo} J.~C.,   {Abu-Saleem} M.,  2022, \mn@doi [Journal of
  High Energy Astrophysics] {10.1016/j.jheap.2022.06.002}, \href
  {https://ui.adsabs.harvard.edu/abs/2022JHEAp..35...83T} {35, 83}

\bibitem[\protect\citeauthoryear{{Zepeda} et~al.,}{{Zepeda}
  et~al.}{2023}]{Zepeda2023}
{Zepeda} J.,  et~al., 2023, \mn@doi [\apj] {10.3847/1538-4357/acbbcc}, \href
  {https://ui.adsabs.harvard.edu/abs/2023ApJ...947...23Z} {947, 23}

\makeatother
\end{thebibliography}

\end{document}